# Evolution of holonic control architectures towards Industry 4.0: A short overview

**Olivier Cardin\*, William Derigent\*\*, Damien Trentesaux\*\*\*.**

\* *LUNAM Université, Université de Nantes, LS2N UMR CNRS 6004*
*2 avenue du Pr Jean Rouxel, 44470 Carquefou, France (e-mail: olivier.cardin@ls2n.fr)*
*\*\* Research Centre for Automatic Control of Nancy, CNRS UMR 7039, Campus Sciences, BP 70239,*
*Vandoeuvre-lès-Nancy Cedex, 54506, France (e-mail: william.derigent@univ-lorraine.fr)*
*\*\*\* LAMIH UMR CNRS 8201, University of Valenciennes and Hainaut-Cambrésis, UVHC*
*Valenciennes, France (e-mail: damien.trentesaux@univ-valenciennes.fr)*

**Abstract:** The flexibility claimed by the next generation production systems induces a deep modification of the behavior and the core itself of the control systems. Overconnectivity and data management abilities targeted by Industry 4.0 paradigm enable the emergence of more flexible and reactive control systems, based on the cooperation of autonomous and connected entities in the decision making process. For the last 20 years, holonic paradigm has become the core paradigm of those evolutions, and evolved in itself. This contribution aims at emphasizing the conceptual evolutions in the application of holonic paradigm in the control architectures of manufacturing systems and highlighting the current research trends in this field.

## 1. INTRODUCTION

The evolution of industrial systems to so-called Industry 4.0 is mainly based on the development of highly connected resources throughout the whole process. This constant flow of information spread by and available for all the actors opens many opportunities to enhance the behavior of the whole process, among which:

- Logistics, with some information transmitted to the whole supply chain to enable a constant synchronization and adaptation to changes;
- Manufacturing, with the perspective to adjust the behavior of the shop floor in real-time due to abnormal conditions or changes;
- Maintenance, with some innovative simulation-based monitoring and knowledge management abilities.

To enable an efficient use of the data available in Industry 4.0 oriented processes, it becomes necessary to adapt the control architecture in order to make it flexible, reactive and adaptable enough to reach the objectives previously described. For the last 20 years, Holonic Control Architectures (HCA) have been widely studied and developed. They happened to be more and more efficient towards those characteristics, and their use at industrial level is starting to spread.

After stating the fundamental vocabulary used in the field of HCA, this article intends to give an overview on the evolution of HCA over the last two decades and how it can contribute to the dissemination of Industry 4.0 technologies.

The current scientific research perspectives are presented in a last section.

## 2. FUNDAMENTALS IN HOLONIC CONTROL ARCHITECTURES

A Holonic Control Architecture (HCA) is an architecture composed of holons, called holarchy. A holon is a communicating decisional entity (with inputs and outputs) composed of a set of sub-level holons and at the same time, part of a wider organization composed of higher-level holons (recursivity, called the Janus effect (Koestler 1978)). It is important to note that a holon is also composed of a physical part associated to a digital one (that can be modeled as a digital agent, avatar, digital twin) and finally, holons are able to decide according to a certain degree of autonomy (Babiceanu and Chen 2006).

Decision making is a huge research topic by itself. In the industrial management community, and from our experience working with HCA, we suggest that deciding is the activity of reducing a set of possibilities. A close notion is the concept of "degree of freedom". From this point of view, classical decision making activities can be derived: choice (set reduced to a singleton), ranking (integrating order in the set), etc. both in discrete (set of resources) and continuous (rotation speed, energy consumption) domains. In HCA, given the importance of physical aspects, deciding is an activity merged into a wider process, that we called a decisional process. Extending the basic ideas of Simon , a decisional process is composed of several activities: monitoring, triggering, design of possible decisions, a priori evaluation of decisions, decision, application and a posteriori evaluation of decisions. Given the recursive aspect of a holarchy, this process is also recursive and can be implemented into layers of holons. For example, the triggering activity for a quality control holon can be decomposed into a decision process handled by lower level holons aiming to decide, through a learning strategy, the best triggering level to avoid over-reaction if too low or to avoid loss of customer if too high.

In this context, **autonomy** is defined as the degree of freedom of each holon regarding its decision capacity, whatever the holon level. It can also be associated to a set of constraints on a search space when using optimization tools. The level of autonomy can be set during the design phase by the designer himself, but it can also be adjusted by a higher level holon with application to a lower level during the exploitation phase. For example, a supervisor holon decides to restrict the set of possible resource-holons to be chosen by lower-level product holons because of a maintenance operation to come on one of these resource-holons. Communication among holons also enables to restrict or to enlarge the autonomy of holons, for example, a direct peer-to-peer negotiation protocol or an indirect through the environment use of pheromone-based communication can lead holons to improve the quality of their decisions through the forbidding of search spaces to avoid local optima during a dynamic task allocation process.

Because of these characteristics of a HCA, it is expected that emerging behavior occurs. From our perspective, an **emerging behavior** is the observation of a property at a higher level of a HCA that has not been explicitly integrated (programmed) into holons composing this HCA. For example, using attractive/repulsive potential field algorithms may lead product-holons to naturally avoid every resource-holon under breakdown and to select them again after recovery without detailing this process in the HCA. As a consequence, several emerging properties can be expected, depending on the objective assigned to the HCA. We denote them as self-* properties.

## 3. ARCHITECTURES EVOLUTION

### 3.1. Benefits and evolution of reference architectures

Implementing a control system based on a set of autonomous communicating entities is a task requiring many skills from the developers. As stated before, several parallel research efforts studied the most efficient control architectures in a general way with a performance point of view, but the development process is not mandatorily enhanced in the same way.

However, considering the development of Holonic Manufacturing Systems, the targeted systems enable to reduce the complexity of the process. Indeed, manufacturing systems generally evolve in a semi-structured environment and handle a set of naturally structured data. As a matter of fact, it is possible to orient and ease the construction of the architecture by specializing a priori some of the holons that will be integrated. The role of the reference architecture is to express the role and the relations between these predefined holons so that the whole architecture could be structured according to the reference.

Fig. 1 introduces the evolution of the HCA definitions along time and highlights the different trends that are currently encountered and their respective impacts on Industry 4.0. Next sections introduce some details about the reference architectures referenced in this evolution.

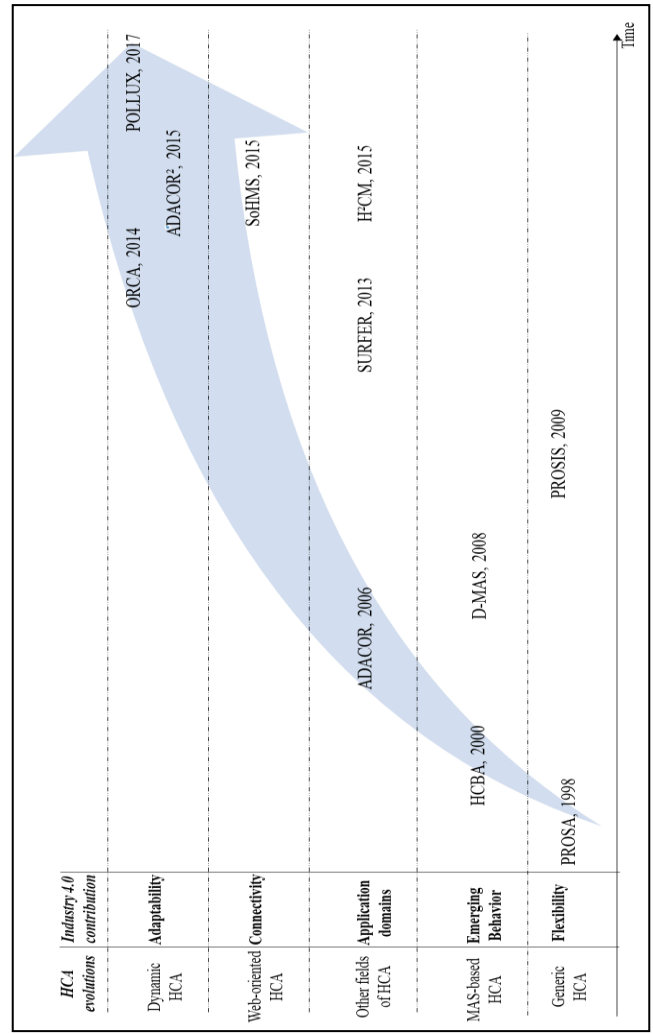

Fig. 1. Evolution of HCA along time and their respective contributions to Industry 4.0

### 3.2. Generic architectures

The first reference architecture that was described in literature is issued from an IMS Project called Holonic Manufacturing Systems in 1996 (Van Brussel et al. 1998). Its acronym is PROSA (Fig. 2), as the acronym of the 4 holons that compose the architecture:

- Product holons, in charge of the management of the production knowledge;
- Resource holons, in charge of the management of the process knowledge;
- Order holons, in charge of the management of the process execution;
- Staff holons, not represented in Fig.2, acting as a global advisor for the whole architecture.

This architecture is the most referenced one in scientific literature (over 1700 times in late 2017) and is often the basis of the emerging architectures as being the most generic one.

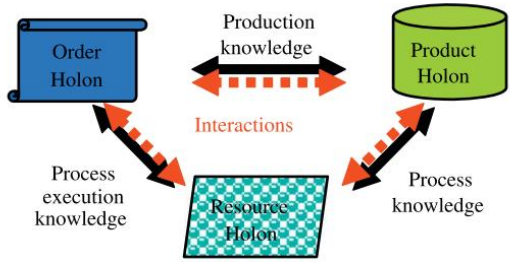

Fig. 2. PROSA simplified architecture

This architecture was the first of a series, developed for different domains (namely manufacturing, logistics, maintenance, etc.). For example, *PROSIS* (Product, Resource, Order, Simulation for Isoarchy Structure) was designed to offer a different organization paradigm using the concept of isoarchy (Pujo et al. 2009). An isoarchy is an architecture containing no subordination hierarchical links between holons. At a same decision level, all the different decision-making entities have exactly the same power in the decision-making mechanism: thus, this corresponds to the total absence of hierarchy, both functionally and temporally. In the PROSIS architecture, the Staff Holon, useless in such isoarchic context, has been replaced by the Simulation Holon which aims to simulate the evolution of the manufacturing system from its current status, obtained via the active listening and analysis of interactions between all other holons.

*3.3. Agent-oriented architectures*

In parallel, the question of implementation of the architectures in a control system perspective was investigated, and several architectures were derived from PROSA in order to benefit from the theories and tools developed in the multi-agent field.

Among them, *HCBA* (Holonic Component Based Architecture) (Chirn and McFarlane 2000) is the first architecture based on a fusion of different concepts originating from component-based development, multi-agent system (MAS) and HMS. The purpose of such a fusion is to develop a highly decentralized architecture, built from autonomous and modular cooperative and intelligent components, able to manage rapidly the different changes, by focusing on the system reconfigurability. HCBA is composed of 2 types of components in the production system, Resource Component and Product Component. The Resource Component is composed of a physical part and a virtual control part. The physical part is dedicated to the execution of operations, whereas its control part manages the execution of operation in the resources, the decision-making process related to resources, and the communication with other components for negotiation. Resources are in charge of the operation scheduling, while looking for the optimization of their use. The Product Component is also composed of a physical part and an informational one. Its physical part can represent materials, parts, pallets and so on. Moreover, the informational part is in charge of the production program, including the routing control, the process control, decision-making and production information. The information part is composed of virtual agents with specific roles. Each Product Component is composed of a Product Coordinator creating WIP agents (Work in Process). Both are in charge of the completion of orders, but at different levels. the Product Coordinator ensures the production monitoring of a lot whereas WIP agents are in charge of the production monitoring of an individual part. As a result, WIP agents negotiate with the resource community to define the part processing in the shop floor. These negotiations are done within an objective set by the Product Coordinator (Fig. 3).

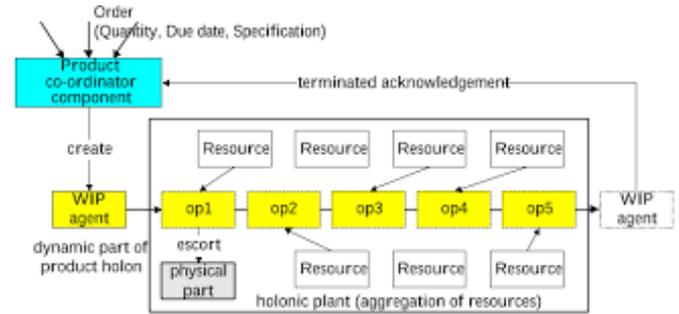

Fig. 3. Structure of the HCBA architecture

In the same way, *Delegate MAS* is an architectural pattern that allows an agent to delegate a responsibility to a swarm of lightweight agents to support this agent in fulfilling its functions. The issuing agent can delegate multiple responsibilities, each of them applying the delegate MAS pattern. The agent may use a combination of delegate multi-agent systems to handle a single responsibility. The delegate MAS may also provide services to other agents. Delegate MAS is a more generic description of an approach inspired by ant food foraging behavior. The delegate MAS pattern consists out of three elements: the agent, the ant and the environment (Fig. 4).

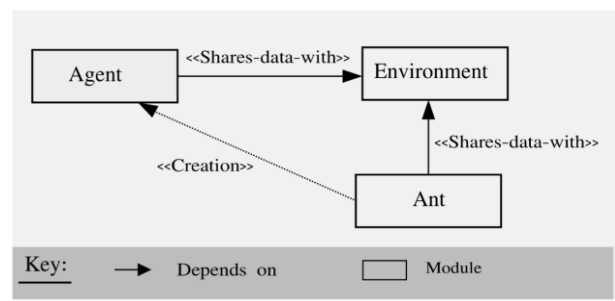

Fig. 4. Delegate MAS architectural view (Verstraete et al. 2008)

*3.4. Extensions of holonic architectures*

*ADACOR* (ADAptive holonic COntrol aRchitecture) is a holonic reference for the distributed manufacturing system, proposed by (Leitão and Restivo 2006). ADACOR is a decentralized control architecture but it also considers centralization in order to tend to a global optimization of the system. Holons are belonging to the following classes: Product Holons (ProdH), Task Holons (TH), Operation

Holons (OpH) and Supervisor Holons (SupH), interconnected via the scheme depicted Fig. 5.

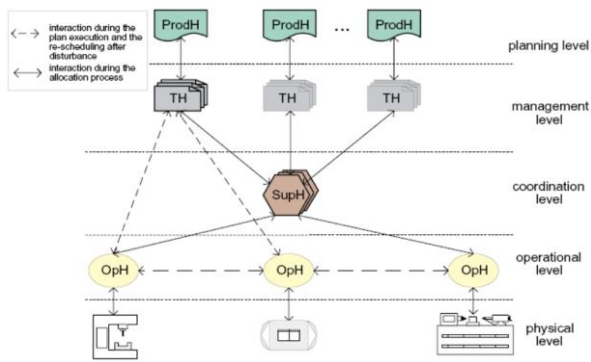

Fig. 5. ADACOR holons repartition (Leitão and Restivo 2006)

SupH are based on biological systems and are different from SH of PROSA. In normal execution, the ADACOR architecture maintains the production system in stationary state, where holons are organized in a hierarchical structure, the OHs following the optimize schedules proposed by the SupHs, for the THs. However, when a manufacturing problem occurs (delay, machine failure, …), the global system enters in transient state, characterized by the re-organization of the holons required to react to the disturbance. To do so, ADACOR uses a pheromone-like spreading mechanism to distribute global information. Thanks to this one, ADACOR introduces the possibility to change dynamically the holarchy between the stationary state and the transient one.

To extend the adaptability of PROSA, an extension to dynamic hybrid systems, i.e. specific systems where both discrete and continuous behaviors need to be taken into account in the control of the system, was presented as $H^2CM$ (Holonic Hybrid Control Model) reference architecture (Indriago et al. 2016). It is based on the three basic holons of PROSA without the Staff one. Two main differences with classical PROSA appear: (i) Each resource is granted with an order and a product along its life. The order holon is in charge of the monitoring of the resource whereas the product holon is in charge of the recipe to be applied on the actual product. The content and objectives of the order and product holons are constantly evolving, but the structure remains permanently the same; (ii) A clear recursivity link is present on the resource holon. Indeed, each compound resource can be fractally decomposed into one or several holarchies, comprising one or several resources and their associated order and product holons. Aggregation relations created here can be changed along the working of the system; holarchies can be created and destroyed online.

SURFER architecture (Le Mortellec et al. 2013) represents an interesting adaptation of control systems for maintenance and monitoring of highly complex systems (namely trains in this case). The generic holonic architectural model proposed for the diagnosis is shown in Fig. 6. This model is composed of recursive diagnosis structures, including sub-systems and their associated diagnosis methods. Each diagnosed system is composed of a control part and a controlled part, which operate in a context. The control part executes an algorithm to control the controlled part and, in return, the controlled part adopts an expected behavior. At the lowest level of the holonic structure, the controlled part is typically composed of physical elements (e.g., sensors, switches and actuators) that are linked to mechanical and electrical constraints.

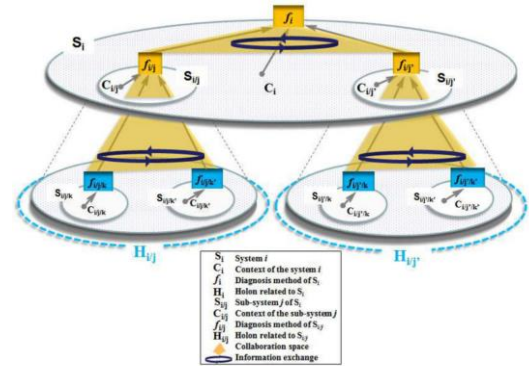

Fig. 6. Basics of the SURFER architecture

### 3.5. Web-oriented HCA

One of the trend in developing HCA is the evolution of cloud-based technologies. In this global trend, a reference HCA was developed in order to cope with some of the web-based standards such as web services. SoHMS architecture (Gamboa Quintanilla et al. 2016) is mainly based on principles and concepts introduced by PROSA (Van Brussel et al. 1998), combining some interaction concepts from HCBA (Chirn and McFarlane 2000) and ADACOR (Barbosa et al. 2015; Leitão and Restivo 2006). This proposal uses the Product (PH), Resource (RH) and Order Holons (OH) from PROSA, and the concept of Directory Facilitator (DF) from multi-agent systems. Even if basic concepts remain close to the original ones, their behavior was adapted for making the services (namely Manufacturing Services – Mservices) the main element of interaction, oriented towards planning and scheduling activities. Fig. 7 shows a class diagram of the architecture detailing the relations and data exchanges between actors. Because of the services perspective, a new element added to the architecture is the SIL (Service Interface Layer), used as an interface between service descriptions and their implementation methods at shop floor. Due to its individual and proprietary character, each resource possesses a SIL instance, containing all the information on the way to implement a service on the lowest level.

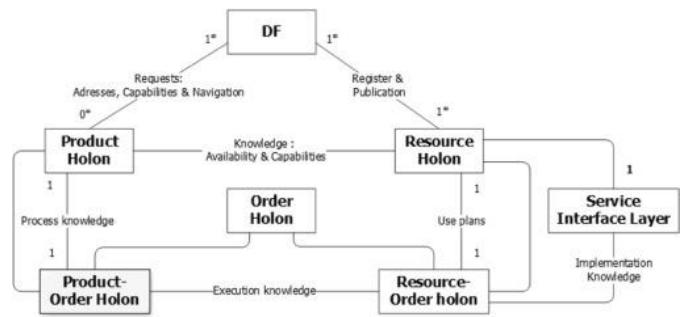

Fig. 7. SoHMS structure

*3.6. Dynamic HCA*

Dynamic HCA definition is probably one of the most promising current trend in literature (Cardin et al. 2015). It postulates that the behavior of the system can be changed dynamically in order to adapt to the changes of the environment and thus reduce the transient states and the associated loss of performance. In the manufacturing domain, ORCA (Pach et al. 2014) was one of the first dynamic HCA that was formalized in literature (Fig. 8).

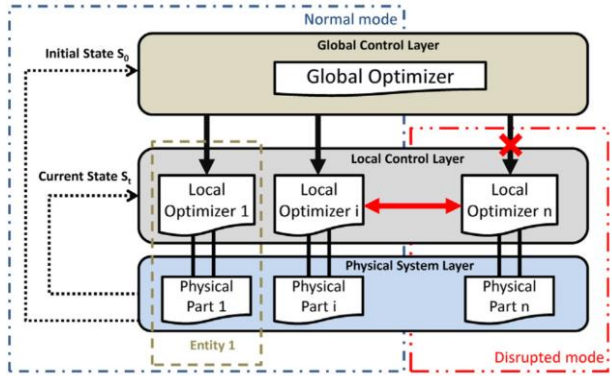

Fig. 8. ORCA global organization (Pach et al. 2014)

An evolution of ADACOR mechanism has also been presented in (Barbosa et al. 2015) as ADACOR². The objective is to let the system evolving dynamically through configurations discovered online, and not only between a stationary and one transient state (Fig. 9). The rest of the architecture is nevertheless quite similar to ADACOR.

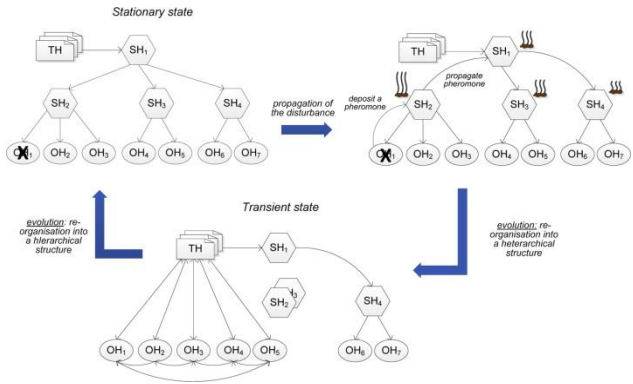

Fig. 9. ADACOR² evolution (Barbosa et al. 2015)

The last architecture in date is denoted as POLLUX (Jimenez et al. 2017). The main novelty is focused on the adaptation mechanism of the architecture, using governance parameters that enlarge or constraint the behavior of the low level holons regarding the disturbances observed by the higher level (Fig. 10), the idea being to find the "best" architecture that suits detected disturbances.

## 4. SCIENTIFIC PERSPECTIVES

During these 20 years of research from the first reference holonic architecture, these ones have evolved, incorporating new features all along time needed to implement the Industry 4.0 vision, as shown Fig. 1. Nevertheless, some additional challenges need to be solved to ensure an easy and secure industrial adoption of such architectures.

*Emerging behavior/adaptability:* in the Industry 4.0 vision, systems should be able to react to unforeseen events, and to propose new behaviors. Current architectures incorporate reaction mechanisms, based on local decisional entities, sometimes coordinated by global decisional entities. One important scientific problem is to correctly adjust the level of autonomy of the local decisional entities against the level of authority of the global decisional entities. Such balance is never easy to achieve and can impact deeply the stability of the manufacturing system.

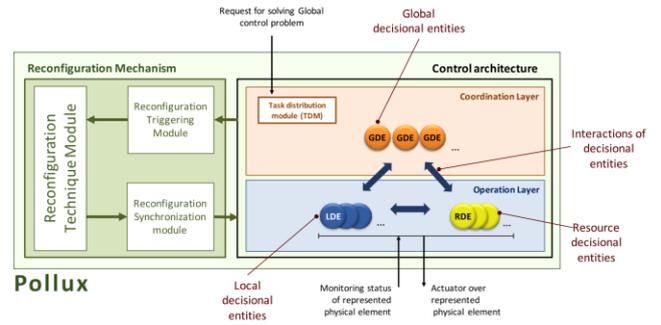

Fig. 10. POLLUX mechanism (Jimenez et al. 2017)

*Myopia of local decisional entities:* local decisional entities are subject to, at least, social and temporal myopia, i.e they suffer from a lack of information concerning their environment and consequences of their actions. This can lead to sub-optimal solutions or even deadlocks. However, myopia is a needed characteristic of holonic architectures, ensuring a fast reaction. Some research papers (Pach et al. 2014) try to address the myopia problem, but a more mathematical definition is needed to completely understand this research problem.

*Performance guarantee:* dynamic HCA like ADACOR², POLLUX or ORCA are interesting, because they integrate an optimal scheduling module used in normal state, coupled with reactive abilities, executed when a disturbance occurs. When it happens, the HCA may modify its own organization to minimize the impact of this disturbance. Such architectures guarantee that performances of the manufacturing system are optimal in normal state, but not always in degraded mode. In fact, because of the level of autonomy and adaptability left to the local decisional entities, ensuring a minimal system performance in every degraded situations could be difficult to attain. This certainly depends on the number and types of disturbances occurring on the manufacturing system. A benchmarking study of each dynamic holonic architecture may then be a very interesting challenge to explore. To guarantee the performance of holonic architectures, formal proofs are another interesting research possibility. This possibility is also interesting with respect to ethical considerations, considering that proofs of behavior are the first steps towards the determination of responsibilities in case of incident. Finally, the generation of the code itself is an important aspect, and a model-driven engineering of the software is probably a response in adequation with the previously presented HCA.

*Sustainability/energy management:* the sustainability issues constitute one of the pillars of Industry 4.0. HCA can be of great benefits for several of those. For example, the management of energy through the notion of smart grids needs a massively distributed, scalable and coordinated control for which a holonic perspective shall be pertinent. The production activity control of the manufacturing or logistics systems will also be impacted by the availability of energy due to renewable energy sources. These kinds of constraints could benefit from a dynamic HCA in order to adapt dynamically to those frequent changes of the environment. A last orientation relies in the connectivity of the consumption elements, which will provide a full access to their real-time and expected consumption for real-time decision making.

*Big Data:* one important aspect of the industry 4.0 is the massive usage of sensors, monitoring the production processes. However, these ones produce an important amount of data that cannot be processed by current HCA. Some pre-processing features should then be added to holonic architectures, to transform data into relevant information for holons, and even into manufacturing knowledge.

*Integration of the human in the loop*: This problem is twofold: first, how to foster the acceptation of these new technologies, capable to take actions with some degree of autonomy? Second, how to define an architecture integrating humans and holons together, taking the best of both?

## 5. SYNTHESIS

This article intends to highlight the evolution of holonic control architectures in the last 20 years in regards to the development of Industry 4.0 paradigm. This study showed that several main trends appeared in the last years in order to cope with the objectives of adaptability and flexibility of the industrial systems. Details were also provided about the available reference HCA in literature, and finally some point of views about the research perspectives that can be expected in this field in the next few years.

This analysis shows a great coherence between HCA evolution and Industry 4.0 objectives, among which:

- The basic concept of HCA which provides autonomy and allows emergent behaviors of the holons fit the Industry 4.0 orientations;
- The extension of the industrial fields of applications of the HCA, in order to integrate all kinds of systems into the Industry 4.0-oriented next generation industrial system;
- The hyper-connectivity of the HCA, notably throughout the cloud-oriented perspective which is meant to gain benefits from the Industry 4.0 based new equipments and requirements for the different machines and organizations;
- The adaptability of industrial systems facing the evolution of their environment is fostered by the dynamic evolution of the HCA, which is still under study but constitutes a promising opportunity.

## REFERENCES

Babiceanu, R. F., & Chen, F. F. (2006). Development and Applications of Holonic Manufacturing Systems: A Survey. Journal of Intelligent Manufacturing, 17(1), 111–131. doi:10.1007/s10845-005-5516-y

Barbosa, J., Leitão, P., Adam, E., & Trentesaux, D. (2015). Dynamic self-organization in holonic multi-agent manufacturing systems: The ADACOR evolution. *Computers in Industry*, *66*, 99–111.

Cardin, O., Trentesaux, D., Thomas, A., Castagna, P., Berger, T., & El-Haouzi, H. B. (2017). Coupling predictive scheduling and reactive control in manufacturing hybrid control architectures: state of the art and future challenges. *Journal of Intelligent Manufacturing*, 28 (7), 1503-1517.

Chirn, J.-L., & McFarlane, D. C. (2000). A holonic component-based approach to reconfigurable manufacturing control architecture. In *11th International Workshop on Database and Expert Systems Applications Proceedings* (pp. 219–223).

Gamboa Quintanilla, F., Cardin, O., L'Anton, A., & Castagna, P. (2016). A modeling framework for manufacturing services in Service-oriented Holonic Manufacturing Systems. *Engineering Applications of Artificial Intelligence*, *55*, 26–36.

Indriago, C., Cardin, O., Rakoto, N., Castagna, P., & Chacòn, E. (2016). H2CM: A holonic architecture for flexible hybrid control systems. *Computers in Industry*, *77*, 15–28.

Jimenez, J.-F., Bekrar, A., Zambrano-Rey, G., Trentesaux, D., & Leitão, P. (2017). Pollux: a dynamic hybrid control architecture for flexible job shop systems. *International Journal of Production Research*, *55*(15), 4229–4247.

Koestler, A. (1978). Janus: A Summing Up. Vintage Books.

Le Mortellec, A., Clarhaut, J., Sallez, Y., Berger, T., & Trentesaux, D. (2013). Embedded holonic fault diagnosis of complex transportation systems. *Engineering Applications of Artificial Intelligence*, *26*(1), 227–240. doi:10.1016/j.engappai.2012.09.008

Leitão, P., & Restivo, F. (2006). ADACOR: A holonic architecture for agile and adaptive manufacturing control. *Computers in Industry*, *57*(2), 121–130.

Pach, C., Berger, T., Bonte, T., & Trentesaux, D. (2014). ORCA-FMS: a dynamic architecture for the optimized and reactive control of flexible manufacturing scheduling. *Computers in Industry*, *65*(4), 706–720.

Pujo, P., Broissin, N., & Ounnar, F. (2009). PROSIS: An isoarchic structure for HMS control. *Engineering Applications of Artificial Intelligence*, *22*(7), 1034–1045.

Van Brussel, H., Wyns, J., Valckenaers, P., Bongaerts, L., & Peeters, P. (1998). Reference architecture for holonic manufacturing systems: PROSA. *Computers in Industry*, *37*(3), 255–274.

Verstraete, P., Saint Germain, B., Valckenaers, P., Van Brussel, H., Van Belle, J., & Karuna, H. (2008). Engineering Manufacturing Control Systems Using PROSA and Delegate MAS. *Int. J. Agent-Oriented Softw. Eng.*, *2*(1), 62–89.